\documentclass[12pt,fleqn]{article}
\usepackage{amsmath}

\begin{document}

\title{\textbf{Cyclic bases of zero-curvature representations:
further examples}}

\author{\textsc{S.~Yu.~Sakovich\footnote{Home institution:
Institute of Physics, National Academy of Sciences, 220072
Minsk, Republic of Belarus. Contact e-mail: saks@pisem.net}}%
\bigskip\\{\footnotesize Mathematical Institute, Silesian
University, 74601 Opava, Czech Republic}}

\date{}

\maketitle

\begin{abstract}
The paper continues nlin.SI/0212019 by giving three more
examples of using cyclic bases of zero-curvature representations
in studies of relation between strong Lax pairs and recursion
operators.
\end{abstract}

\section{Introduction}

The present paper continues \cite{Sak1} and provides further
examples of using cyclic bases of zero-curvature representations.

A zero-curvature representation (ZCR) of a system of differential
equations is the compatibility condition
\begin{equation}
\label{e1}
Z \equiv D_t X - D_x T + [ X , T ] = 0
\end{equation}
of the overdetermined linear system
\begin{align}
\Psi_x & = X \Psi , \label{e2} \\
\Psi_t & = T \Psi , \label{e3}
\end{align}
where $D_x$ and $D_t$ denote the total derivatives; $X$ and $T$
are $n \times n$ matrix functions of the independent variables
$x$ and $t$, dependent variables $u^i(x,t)$ and finite-order
derivatives of $u^i$; $\Psi (x,t)$ is an $n$-component column;
the square brackets stand for the matrix commutator; and the
condition \eqref{e1} is satisfied by any solution $u^i$ of the
represented system. Any linear transformation
\begin{equation}
\label{e4}
\Psi'(x,t) = G \left( x, t, u^i, u_x^i, \dotsc
\right) \Psi (x,t), \qquad \det G \neq 0
\end{equation}
generates the gauge transformation
\begin{equation}
\label{e5}
X' = G X G^{-1} + ( D_x G ) G^{-1}, \qquad
T' = G T G^{-1} + ( D_t G ) G^{-1},
\end{equation}
which results in the equivalence transformation
$Z' = G Z G^{-1}$ of \eqref{e1}. For this reason, ZSRs should
be studied by some gauge-invariant techniques. Gauge-invariant
formulations of ZCRs were proposed in \cite{Mar1,Mar2} and
\cite{Sak2}, independently.

The concept of cyclic bases of ZCRs was introduced in
\cite{Sak2}. It proved to be useful for distinguishing between
`good' and `bad' Lax pairs, checking if parameters in ZCRs are
essential, constructing complete hierarchies associated with
given linear problems, deriving recursion operators, and so on
\cite{Sak1,Sak2,Sak3,Sak4,KKS,Sak5,KSY,Sak6}. Sections \ref{s2},
\ref{s3} and \ref{s4} of the present paper provide further
illustrations to this concept.

\section{AKNS-related spectral problem} \label{s2}

Let us consider the linear equation \eqref{e2} with
\begin{equation}
\label{e6}
X =
\begin{pmatrix}
\frac{1}{2} u - \frac{1}{2} \lambda & - v \\
v & - \frac{1}{2} u + \frac{1}{2} \lambda
\end{pmatrix}
,
\end{equation}
where $u$ and $v$ are functions of $x$ and $t$, and $\lambda$ is
a parameter. This spectral problem, recently studied in
\cite{Chen}, is hardly new, because \eqref{e5} with a diagonal
matrix $G$ can bring it into the AKNS form. Anyhow, we take this
spectral problem `as is' and use it to show how the cyclic basis
method automatically generates a recursion operator for the
hierarchy studied in \cite{Chen}.

Following the way described in \cite{Sak1,Sak2,Sak4,KKS,Sak6},
we pose the problem of finding all systems of evolution equations
\begin{equation}
\label{e7}
\begin{split}
u_t & = f ( x, t, u, v, \dotsc ,
u_{x \dotsc x} , v_{x \dotsc x} ) , \\
v_t & = g ( x, t, u, v, \dotsc ,
u_{x \dotsc x} , v_{x \dotsc x} )
\end{split}
\end{equation}
which admit ZCRs \eqref{e1} with $X$ given by \eqref{e6}. The
characteristic matrices are $C_u = \partial X / \partial u$ and
$C_v = \partial X / \partial v$ because $X$ \eqref{e6} contains
no derivatives. Applying the operator $\nabla$, defined by
$\nabla Y = D_x Y - [ X , Y ]$ for any (here, $2 \times 2$) matrix
$Y$, to $C_u$, $C_v$ and $\nabla C_u$, we find the cyclic basis to
be three-dimensional, $\{ C_u , C_v , \nabla C_u \}$, with the
closure equations
\begin{equation}
\label{e8}
\begin{split}
& \nabla C_v = - ( u - \lambda ) v^{-1} \nabla C_u , \\
& \nabla^2 C_u = - 4 v^2 C_u - ( u - \lambda ) v C_v +
v^{-1} v_x \nabla C_u ,
\end{split}
\end{equation}
where the dependence of gauge-invariant coefficients on $\lambda$
proves that $\lambda$ is an essential parameter in $X$ \eqref{e6}.

Decomposing $T$ as
\begin{equation}
\label{e9}
T = p C_u + q C_v + r \nabla C_u ,
\end{equation}
where $p$, $q$ and $r$ are any functions of $\lambda, x, t, u, v,
\dotsc , u_{x \dotsc x} , v_{x \dotsc x}$, and using the
characteristic form
\begin{equation}
\label{e10}
f C_u + g C_v = \nabla T
\end{equation}
of \eqref{e1} and the closure equations \eqref{e8}, we find that
\begin{gather}
p = ( u - \lambda ) v^{-1} q - D_x r - v^{-1} v_x r
\label{e11} , \\
h = ( M - \lambda N ) s: \qquad
h =
\begin{pmatrix}
f \\ g
\end{pmatrix}
, \qquad
s =
\begin{pmatrix}
q \\ r
\end{pmatrix}
\label{e12} ,
\end{gather}
where
\begin{gather}
M =
\begin{pmatrix}
D_x \circ u v^{-1} & - D_x^2 - D_x \circ v^{-1} v_x - 4 v^2 \\
D_x & - u v
\end{pmatrix}
\label{e13} , \\
N =
\begin{pmatrix}
D_x \circ v^{-1} & 0 \\
0 & - v
\end{pmatrix}
\label{e14} .
\end{gather}

Then, using the expansion
\begin{equation}
\label{e15}
s = s_0 + \lambda s_1 + \lambda^2 s_2 + \dotsb
\end{equation}
and taking into account the condition
\begin{equation}
\label{e16}
\partial h / \partial \lambda = 0 ,
\end{equation}
we get from \eqref{e12} the following solution of the problem:
a system \eqref{e7} admits a ZCR \eqref{e1} with $X$ \eqref{e6}
iff its right-hand side $h$ is determined by
\begin{equation}
\label{e17}
h = M s_0 ,
\end{equation}
where $s_0$ belongs to a set of functions $s_0,s_1,s_2,\dotsc$
satisfying the recurrence
\begin{equation}
\label{e18}
N s_i = M s_{i+1}, \qquad i = 0, 1, 2, \dotsc .
\end{equation}

The set $\{ s'_i \} : s'_i = N^{-1} M s_i$ satisfies \eqref{e18}
if a set $\{ s_i \}$ does, therefore $h' = M N^{-1} h$ is the
right-hand side of a represented system if $h$ is, and
\begin{equation}
\label{e19}
R = M N^{-1}
\end{equation}
is a recursion operator for the represented hierarchy. Inverting
$N$ \eqref{e14},
\begin{equation}
\label{e20}
N^{-1} =
\begin{pmatrix}
v D_x^{-1} & 0 \\
0 & - v^{-1}
\end{pmatrix}
,
\end{equation}
we obtain the recursion operator \eqref{e19} explicitly:
\begin{equation}
\label{e21}
R =
\begin{pmatrix}
D_x \circ u D_x^{-1} & D_x \circ v^{-1} D_x + 4 v \\
D_x \circ v D_x^{-1} & u
\end{pmatrix}
.
\end{equation}

\section{ZCR of a linear equation} \label{s3}

Our next example is a ZCR of the linear equation
\begin{equation}
\label{e22}
u_t = u_{xx} ,
\end{equation}
namely, \eqref{e1} with
\begin{gather}
X =
\begin{pmatrix}
\alpha & u \\
0 & - \alpha
\end{pmatrix}
, \label{e23} \\
T =
\begin{pmatrix}
2 \alpha^2 & u_x + 2 \alpha u \\
0 & - 2 \alpha^2
\end{pmatrix}
, \label{e24}
\end{gather}
where $\alpha$ is a parameter. The problem is to derive the
evident recursion operator $R=D_x$ of \eqref{e22} from this ZCR.

From $X$ \eqref{e23}, we find the characteristic matrix and its
first covariant derivative to be
\begin{equation}
\label{e25}
C =
\begin{pmatrix}
0 & 1 \\ 0 & 0
\end{pmatrix}
, \qquad
\nabla C = - 2 \alpha C .
\end{equation}
We see that the cyclic basis is one-dimensional in this case,
consisting of $C$ only, and that $\alpha$ is an essential
parameter, not removable by gauge transformations. Also we see
that the matrix $T$ \eqref{e24} cannot be decomposed over the
cyclic basis only. For this reason, we have to take into account
the singular basis, a necessary extension of the cyclic basis
\cite{Sak2}:
\begin{equation}
\label{e26}
S =
\begin{pmatrix}
1 & 0 \\ 0 & - 1
\end{pmatrix}
: \qquad
\nabla S = 2 u C .
\end{equation}

Now, the matrix $T$ \eqref{e24} can be decomposed over $\{C,S\}$,
and we pose the problem of finding all evolution equations
\begin{equation}
\label{e27}
u_t = f ( x, t, u, u_x, \dotsc , u_{x \dotsc x} )
\end{equation}
which admit ZCRs \eqref{e1} with $X$ given by \eqref{e23} and $T$
of the form
\begin{equation}
\label{e28}
T = p C + \sigma S ,
\end{equation}
where $p$ and $\sigma$ are any functions of $\alpha, x, t, u,
\dotsc, u_{x \dotsc x}$. Using the characteristic form
\begin{equation}
\label{e29}
f C = \nabla T
\end{equation}
of \eqref{e1} and the closure equations \eqref{e25} and
\eqref{e26}, we find that
\begin{gather}
D_x \sigma = 0, \label{e30} \\
f = ( M - \lambda N ) p + 2 \sigma u , \label{e31}
\end{gather}
where
\begin{equation}
\label{e32}
M = D_x , \qquad N = 1 , \qquad \lambda = 2 \alpha .
\end{equation}

Though \eqref{e31} is similar to \eqref{e12}, the additional term
$2 \sigma u$, which appeared in the right-hand side of \eqref{e31}
owing to the presence of the singular basis, does not allow us to
conclude immediately that the represented hierarchy admits the
recursion operator
\begin{equation}
\label{e33}
R = M N^{-1} .
\end{equation}
Nevertheless, the result \eqref{e33} is correct. Let us make use
of the condition \eqref{e30}. Applying the operator
\begin{equation}
\label{e34}
Q = D_x \circ u^{-1}
\end{equation}
to the left-hand and right-hand sides of \eqref{e31}, we obtain
\begin{equation}
\label{e35}
g = \left( \widetilde{M} - \lambda \widetilde{N} \right) p ,
\end{equation}
where
\begin{equation}
\label{e36}
g = Q f, \qquad \widetilde{M} = Q M, \qquad \widetilde{N} = Q N.
\end{equation}
Then, using the expansion $p = \sum_{i=0}^{\infty} \lambda^i p_i
( x, t, u, \dotsc , u_{x \dotsc x} )$ and taking into account the
condition $\partial g / \partial \lambda = 0$ which follows from
$\partial f / \partial \lambda = 0$, we conclude that $g' =
\widetilde{M} \widetilde{N}^{-1} g$ satisfies \eqref{e35} with
$p' = \widetilde{N}^{-1} \widetilde{M} p$ if $g$ does with some
$p$. Finally, owing to \eqref{e36}, we have $f' = Q^{-1} g' =
M N^{-1} Q^{-1} g = M N^{-1} f$, which proves \eqref{e33} and
leads to the expected $R = D_x$.

\section{Ibragimov--Shabat system} \label{s4}

The last example in this paper is the system
\begin{equation}
\label{e37}
u_t = u_{xx} + \tfrac{1}{2} v^2 , \qquad v_t = 2 v_{xx} ,
\end{equation}
introduced in \cite{IS}. For a long time, this system was thought
to possess only one local generalized symmetry, namely, the
third-order one: see e.g.\ \cite{IS} itself and the first edition
of \cite{Olv}. Later, in the second edition of \cite{Olv}, in
Exercise 5.16(a), it was pointed out that \eqref{e37} really
possesses local generalized symmetries of higher orders, and it
was proposed to find a higher order symmetry and a recursion
operator.

According to the recent classification \cite{BSW}, the system
\eqref{e37} belongs to one of the nine exceptional cases of
Bakirov's systems (it is the $\mathcal{B}_2 [ \tfrac{1}{2} ]$
case) and possesses infinitely many local generalized symmetries
therefore. Moreover, \cite{BSW} contains a recurrent procedure
which allows to construct a symmetry of order $n$ for \eqref{e37}
from its two symmetries of orders $n-1$ and $n-2$. This recurrent
procedure is, however, not a recursion operator, in the sense that
a recursion operator should produce one symmetry from one symmetry.

We can easily find a formal recursion operator $R$ of \eqref{e37},
or a `formal symmetry', which satisfies the condition $D_t (R) =
[ F , R ]$, where $F$ is the Fr\'echet derivative of the
right-hand side of the system \eqref{e37}. The result is
\begin{gather}
R =
\begin{pmatrix}
D_x & d \\ 0 & 2 D_x
\end{pmatrix}
, \label{e38} \\
d = v D_x^{-1} + v_x D_x^{-2} + v_{xx} D_x^{-3} +
v_{xxx} D_x^{-4} + \dotsb . \label{e39}
\end{gather}
According to \cite{Serg}, such formal recursion operators do exist
for wide classes of block-triangular evolution systems, including
all Bakirov's systems. It is, however, not clear, how to apply the
formal expansion \eqref{e39} to local expressions, or how to bring
the formal recursion operator \eqref{e38}--\eqref{e39} into some
`closed' form.

We do not believe that the system \eqref{e37} possesses any
`usual' recursion operator, which can be written in the quotient
form
\begin{equation}
\label{e40}
R = M N^{-1} ,
\end{equation}
where $M$ and $N$ are linear matrix differential operators with
local coefficients. Let us show that the cyclic basis method,
which---as far as we know---works well for systems possessing
recursion operators of the quotient form \eqref{e40}, does not
produce any recursion operator in the case of \eqref{e37}.

It is an easy exercise to find that the Ibragimov--Shabat system
\eqref{e37} admits the ZCR \eqref{e1} with
\begin{gather}
X =
\begin{pmatrix}
\lambda & v & 0 & - 8 u \\
0 & 0 & 1 & 0 \\
0 & 0 & 0 & v \\
0 & 0 & 0 & - \lambda
\end{pmatrix}
, \label{e41} \\
T =
\begin{pmatrix}
2 \lambda^2 & 2 \lambda v + 2 v_x & - 2 v &
- 16 \lambda u - 8 u_x \\
0 & 0 & 4 \lambda & 2 v \\
0 & 0 & 0 & 2 \lambda v + 2 v_x \\
0 & 0 & 0 & - 2 \lambda^2
\end{pmatrix}
, \label{e42}
\end{gather}
where $\lambda$ is a parameter. Then we find from $X$ \eqref{e41}
that the cyclic basis is three-dimensional, $\{ C_u , C_v ,
\nabla C_v \}$, with the closure equations
\begin{equation}
\label{e43}
\nabla C_u = - 2 \lambda C_u , \qquad
\nabla^2 C_v = - \tfrac{1}{4} v C_u - \lambda^2 C_v
- 2 \lambda \nabla C_v
\end{equation}
which clearly show that $\lambda$ is an essential parameter. Since
the matrix $T$ \eqref{e42} cannot be decomposed over the cyclic
basis only, we add the two-dimensional singular basis
$\{ S_1, S_2 \}$ with the elements
\begin{equation}
\label{e44}
S_1 =
\begin{pmatrix}
1 & 0 & 0 & 0 \\
0 & 0 & 0 & 0 \\
0 & 0 & 0 & 0 \\
0 & 0 & 0 & - 1
\end{pmatrix}
, \qquad
S_2 =
\begin{pmatrix}
0 & 0 & 0 & 0 \\
0 & 0 & 1 & 0 \\
0 & 0 & 0 & 0 \\
0 & 0 & 0 & 0
\end{pmatrix}
\end{equation}
which satisfy the closure equations
\begin{equation}
\label{e45}
\nabla S_1 = 2 u C_u + v C_v , \qquad
\nabla S_2 = - \lambda v C_v - v \nabla C_v .
\end{equation}

Now, $T$ \eqref{e42} can be decomposed as
\begin{equation}
\label{e46}
T = ( u_x + 2 \lambda u ) C_u + 2 v_x C_v - 2 v \nabla C_v
+ 2 \lambda^2 S_1 + 4 \lambda S_2 ,
\end{equation}
and we pose the problem of finding all systems
\begin{equation}
\label{e47}
\begin{split}
u_t & = f ( x, t, u, v, \dotsc ,
u_{x \dotsc x} , v_{x \dotsc x} ) , \\
v_t & = g ( x, t, u, v, \dotsc ,
u_{x \dotsc x} , v_{x \dotsc x} )
\end{split}
\end{equation}
which admit ZCRs \eqref{e1} with $X$ given by \eqref{e41} and $T$
of the form
\begin{equation}
\label{e48}
T = p C_u + q C_v + r \nabla C_v + \sigma S_1 + \tau S_2 .
\end{equation}
The characteristic form
\begin{equation}
\label{e49}
f C_u + g C_v = \nabla T
\end{equation}
of these ZCRs leads through the closure equations \eqref{e43} and
\eqref{e45} to
\begin{gather}
D_x \sigma = D_x \tau = 0, \label{e50} \\
q = - D_x r + 2 \lambda r + \tau v \label{e51}
\end{gather}
and
\begin{equation}
\label{e52}
\begin{split}
f & = D_x p - 2 \lambda p - \tfrac{1}{4} v r + 2 \sigma u , \\
g & = - D_x^2 r + 2 \lambda D_x r - \lambda^2 r + \sigma v +
\tau ( v_x - \lambda v ) .
\end{split}
\end{equation}

Conditions \eqref{e50} allow us to remove from \eqref{e52} the
three terms caused by the presence of the singular basis, and we
obtain
\begin{equation}
\label{e53}
h = \left( \widetilde{M} + \lambda \widetilde{L}
+ \lambda^2 \widetilde{K} \right) s ,
\end{equation}
where
\begin{equation}
\label{e54}
\begin{split}
& h = Q
\begin{pmatrix}
f \\ g
\end{pmatrix}
, \qquad
s =
\begin{pmatrix}
p \\ r
\end{pmatrix}
, \\
& \widetilde{M} = Q M , \qquad
\widetilde{L} = Q L , \qquad
\widetilde{K} = Q K ,
\end{split}
\end{equation}
with
\begin{equation}
\label{e55}
M =
\begin{pmatrix}
D_x & - \frac{1}{4} v \\ 0 & - D_x^2
\end{pmatrix}
, \qquad
L =
\begin{pmatrix}
- 2 & 0 \\ 0 & 2 D_x
\end{pmatrix}
, \qquad
K =
\begin{pmatrix}
0 & 0 \\ 0 & - 1
\end{pmatrix}
\end{equation}
and
\begin{equation}
\label{e56}
Q =
\begin{pmatrix}
D_x \circ u^{-1} & 0 \\
0 & D_x \circ \bigl( \left( v^{-1} v_x \right)_x \bigr)^{-1}
D_x \circ v^{-1}
\end{pmatrix}
.
\end{equation}
The operator $Q$ \eqref{e56} does not contain $\lambda$,
therefore the condition $\partial h / \partial \lambda = 0$
follows from $\partial f / \partial \lambda = \partial g /
\partial \lambda = 0$. Taking this into account and substituting
\begin{equation}
\label{e57}
s = s_0 + \lambda s_1 + \lambda^2 s_2 + \dotsb
\end{equation}
into \eqref{e53}, we find that
\begin{equation}
\label{e58}
h = \widetilde{M} s_0
\end{equation}
and that the set of two-component functions $s_i (x,t,u,v,\dotsc,
u_{x \dotsc x},v_{x \dotsc x})$ must satisfy the relations
\begin{equation}
\label{e59}
s_1 = A s_0 , \qquad s_{i+2} = A s_{i+1} + B s_i ,
\qquad i = 0, 1, 2, \dotsc ,
\end{equation}
where
\begin{equation}
\label{e60}
\begin{split}
& A = - M^{-1} L =
\begin{pmatrix}
2 D_x^{-1} & \frac{1}{2} D_x^{-1} \circ v D_x^{-1} \\
0 & 2 D_x^{-1}
\end{pmatrix}
, \\
& B = - M^{-1} K =
\begin{pmatrix}
0 & - \frac{1}{4} D_x^{-1} \circ v D_x^{-2} \\
0 & - D_x^{-2}
\end{pmatrix}
.
\end{split}
\end{equation}

The operators \eqref{e60} are so simple that we can solve the
recurrence \eqref{e59} explicitly and obtain the following:
\begin{equation}
\label{e61}
s_{i+1} =
\begin{pmatrix}
2 D_x^{-1} & \frac{i+2}{4i+4} D_x^{-1} \circ v D_x^{-1} \\
0 & \frac{i+2}{i+1} D_x^{-1}
\end{pmatrix}
s_i , \qquad i = 0, 1, 2, \dotsc .
\end{equation}
We see from \eqref{e61} that, in contrast to what was observed for
all ZCRs studied by the cyclic basis method before, these $s_i$
are not related with each other as $s_{i+1} = P s_i$ with some
operator $P$ independent of $i$, and we get no recursion operator
therefore. The origin of this phenomenon is twofold. First, the
quadratic dependence on $\lambda$ of the matrix differential
operator in \eqref{e53} leads to the three-term recurrence
\eqref{e59}, whereas a linear dependence on $\lambda$ always leads
to a two-term recurrence which produces a recursion operator
automatically. Second, though such a quadratic dependence on the
spectral parameter was also observed in \cite{KKS}, a recursion
operator appeared there owing to the specific condition
$K L^{-1} K = 0$, which is not satisfied by the operators $K$ and
$L$ \eqref{e55} in the present case.

\section*{Acknowledgments}

The author is grateful to the Mathematical Institute of Silesian
University for hospitality, and to the Centre for Higher
Education Studies of Czech Republic for support.

\end{document}